# Pulse generation with ultra-superluminal pulse propagation in semiconductor heterostructures by superradiant phase transition enhanced by transient coherent population gratings

Peter P. Vasil'ev[1,2*], Richard V. Penty[1], and Ian H. White[1]

[1] Centre for Photonic Systems, Department of Engineering, University of Cambridge, 9 JJ Thomson Avenue, Cambridge, CB3 0FA, UK

[2] Quantum Electronics Division, PN Lebedev Physical Institute, 53 Leninsky Prospect, Moscow 119991, Russia

*Email: pv261@cam.ac.uk

*This paper reports the observation of ultra-superluminal pulse propagation in GaAs/AlGaAs multiple contact heterostuctures in a superradiant emission regime, and shows definitively that it is a different class of emission from conventional spontaneous or stimulated emission. It is shown that coherent population gratings induced in the semiconductor medium under strong electrical pumping have great impact in causing a decrease of the group refractive index in the range of 5-40%. This decrease is much greater than that which would be observed due to conventional carrier depletion or chirp mechanisms. The decrease in refractive index in turn causes faster-than-c propagation of femtosecond pulses. The measurement also shows unequivocally the exist of coherent amplification of electromagnetic pulses in semiconductors at room temperature, the coherence being strongly enhanced by interactions of the light with coherent transient gratings locked to carrier gratings. This pulse generation technique can be anticipated to have great potential in applications where highly coherent femtosecond optical pulses must be generated on demand.*

## I. Introduction

The study of cooperative emission from an ensemble of quantum oscillators, often referred as superradiance (SR) or superfluorescence, was triggered by a pioneering paper of Dicke in 1954 [1]. Since then exciting theoretical and experimental research has concerned the collective quantum behaviour and SR emission from different types of media, including gases, solids, polymers, Bose condensates, and 0D, 2D, and 3D semiconductors [2-8]. The model describing the interaction of matter with electromagnetic radiation proposed by Dicke has proven to be of a key importance for studying collective, coherent, and dynamical effects in quantum optics [9]. The real interaction between electromagnetic emission and matter is too complicated for a complete theoretical investigation. However, Dicke found how to make a significant simplification which allowed him to consider physical phenomena within the framework of a simple analytic model which has exact solutions. Here it is assumed that the system is composed of $N$ emitters cooperatively interacting with a single radiation field mode. The essence of cooperative behaviour is that the oscillator dipoles interact coherently with the privileged radiation mode. One of the most striking phenomena provided by the Dicke model is the ability of the system to exhibit a second order phase transition from a normal to a superradiant state at a certain critical temperature [10-12]. In recent years, the Dicke model has received renewed interest because it is a simple system in which one can find entanglement and related phenomena, and because it can be realized in systems more widely than in the original cavity quantum electrodynamics case. A new aspect emerged when it was realized that the SR quantum phase transition is relevant to quantum information and quantum computing [13].

One of the most fascinating and important phenomena which occur during an SR phase transition is the self-organization of emitters and the formation of a macroscopically ordered state. The mutual phasing of emitters involved in radiative emission, their self-organisation and self-order originates from an exchange of photons of internal electromagnetic field. Strong pumping and large field-matter coupling are the crucial factors which lead to a discontinuous change of the system from a normal to a superradiant state. For example, the experimental observations of Dicke quantum phase transition in Bose condensates of ultracold atoms [14] and microcavity polaritons at cryogenic temperatures [15] requires high-finesse optical cavities and strong pumping the system above the critical level.  The self-organization of emitters brings about regular 1D (grating-like) or 2D (chequerboard-like) spatial distributions of the inverted populations. Likewise, two counterpropagating SR pulses burn a periodic modulation into the inverted population distributions, the period of the modulation being half the optical wavelength. This lattice or grating results in a backscattering and pulse correlation [16,17]. As known from classical theory, population grating leads to refractive index gratings in semiconductor materials and can result in a change in group velocity of light. As "fast" and "slow" light have been recently attracted a great deal of interest within the physics community [18-21], the control of the speed of the propagation of light, as demonstrated here, could potentially find a number of applications in optical communications, photonics, optical storage, and other fields of quantum electronics.

Various papers have reported SR phenomena in semiconductors at room temperature [22-25]. Such observations however have led to debate on whether the pulses generated are really akin to those superradiant phenomena observed in low temperature or low density structures [5-7], and whether they are different, for example, than those generated by conventional Q-switching [26]. In this work, therefore, by measuring the superluminal propagation properties of the pulses generated in GaAs/AlGaAs multiple contact heterostructures operating in a superradiant emission regime, we demonstrate for the first time the clear existence of transient coherent gratings directly coupled to the amplified pulses. We observe a decrease of the group refractive index in the range of 5-40%, causing a strong superluminal behaviour not seen in conventional spontaneous or stimulated processes, and due to the SR phase transition. We present experimental and theoretical results of the superluminal pulse propagation brought about during the SR phase transition in the semiconductor medium at room temperature due to coherent transient refractive index gratings.

**II. Experiment**

In the experimental work, a large variety of multiple section GaAs/AlGaAs bulk laser structures capable of generating SR emission have been studied. In general, they have gain and saturable absorber sections of different geometries with different gain/absorber ratios and total cavity lengths. The devices under test are described in detail in our previous publications (see, for instance, [22-25]). The composition of the GaAs/AlGaAs heterostructures is varied. This results in a broad range of the operating wavelengths between 820 and 890 nm. All devices are able to operate under continuous wave operation, gain/Q-switching or SR regimes depending on the driving conditions [26]. A typical 3-section laser structure is schematically illustrated in Figure 1. The two end sections are pumped by nanosecond current pulses providing two areas with high e-h density. The centre section is reverse biased and is a controllable saturable absorber. Generated pulses can travel back and forth between the chip facets whose power reflectivity coefficients are around 0.32.

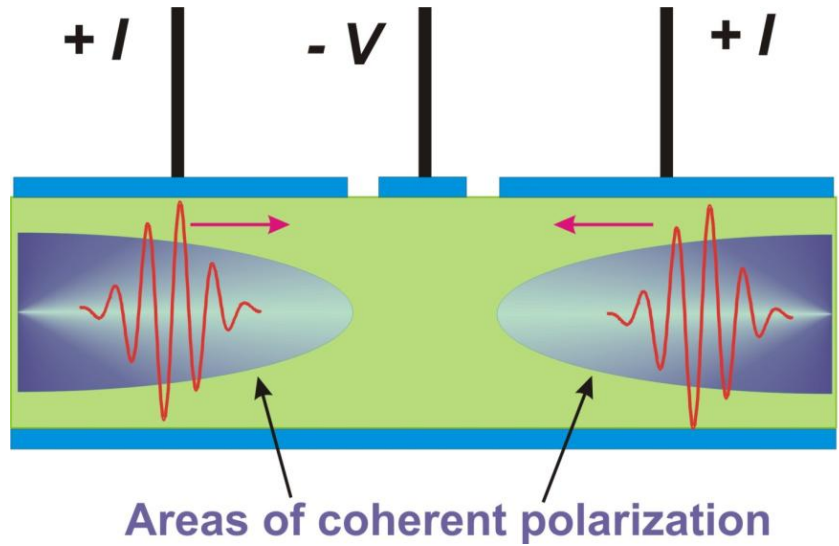

Figure 1. Schematic of the semiconductor structure. The forward driving current *I* and the reverse bias *V* are applied to the end and central sections, respectively.

The regions of coherent polarized amplification (the macroscopically ordered state of the e-h system) exist at both ends of the structure during the superradiant phase transition, as these are the regions that are strongly pumped [22,23]. The exact values of the cavity lengths of devices are measured using SEM with an accuracy of better than 1 micron, as accurate values are required for the calculation of the group refractive index.

We calculate variations of the group refractive index using measurements of both longitudinal mode spectral spacings and round trip times of pulses in different dynamic regimes. Optical emission spectra are analyzed by an optical spectrum analyzer with a spectral resolution of 0.07 nm. The round trip time is measured with femtosecond accuracy by a fringe-resolved autocorrelation technique based on second harmonic generation (SHG) [26] or using a single shot streak camera with an ultimate temporal resolution of about 1.5 ps.

Figure 2 presents typical optical spectra of standard c.w. lasing and SR emission generated from the same 3-section structure at different reverse bias levels and gain currents. SR spectra are always red shifted by 10-20 nm with respect to the peak of lasing as the observed cooperative emission occur at the band gap energy [22-25]. The peak wavelength of the SR emission is around 890 nm, whereas the centre wavelength of the c.w. emission is located at 877 nm.

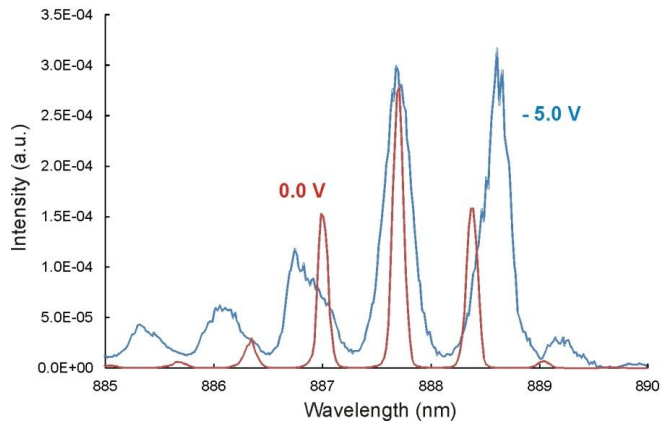

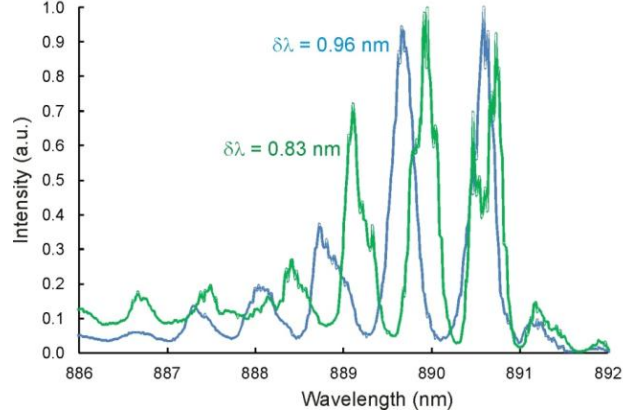

Figure 2. Top - c.w. lasing emission spectrum (red line) and SR spectrum (blue line) of a 100 μm long 3-section structure. Bottom – two SR optical spectra at I = 380 mA, V= - 4.9 V (green line) and I = 450 mA, V= - 4.4 V (blue line)

For better understanding, the c.w spectrum in Fig. 2 (top) is plotted red-shifted by about 13 nm and is seen to have a different mode spacing. As expected, the measured mode spacing of the lasing spectra is dependent on the cavity length but does not depend on driving conditions. By contrast, SR spectra exhibit a clear dependence of the spacing on current and voltage.

We have experimentally observed that samples with different cavity lengths exhibited different refractive index decreases. Figure 3 presents the experimentally measured changes to the mode spacing and the group refractive index with increasing voltage $V$ to the absorber section, this having a 150 μm length. The group refractive index is calculated according to the relation $n_g = \lambda_0^2 / 2L\Delta\lambda$ where $\lambda_0$ is the central wavelength, $L$ is the chip length, and $\Delta\lambda$ is the mode spacing.

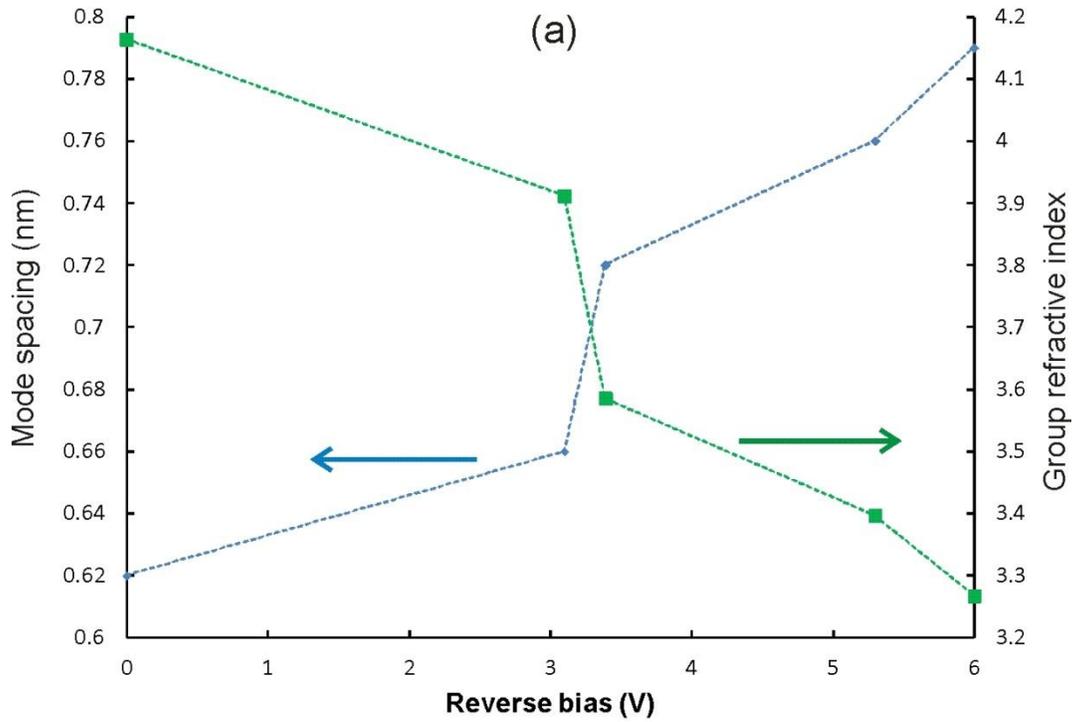

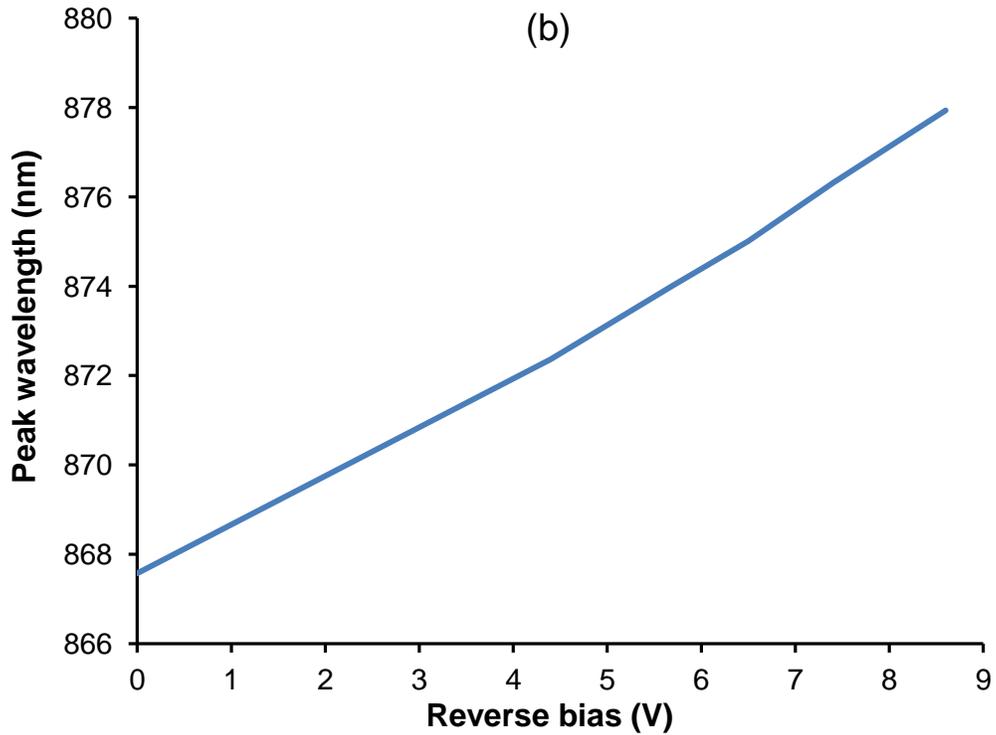

Figure 3. A dependence of the SR mode spacing and corresponding change of the group refraction index (a) and peak wavelength (b) on the reverse bias $V$.

Figure 4 shows a typical burst of SR pulses from a 450 μm long device. There are 2 pulses on the period of the resonator. The number of pulses is normally between 2 and 6 depending on driving and the cavity length. The pulse widths of the individual pulses are less than 1 ps. They are accurately measured using the SHG autocorrelation technique [26]. The common feature of all SR pulses is that the separation between the pulses is always smaller than a half of RTT, which is measured for the standard lasing regime in the same semiconductor structure.

Figure 5 illustrates a number of typical fringe-resolved autocorrelation traces of SR pulses. The detection of SHG autocorrelation traces allows for the direct and accurate measurement of the round trip time. It is well-known [26] that a fringe-resolved autocorrelation trace of any laser emission exhibits a *single* peak at zero delay. Its width determines the coherence time of the laser emission and is inversely proportional to the spectral bandwidth. For mode-locked pulses there exist additional *identical* peaks separated by the round trip time of the laser cavity.

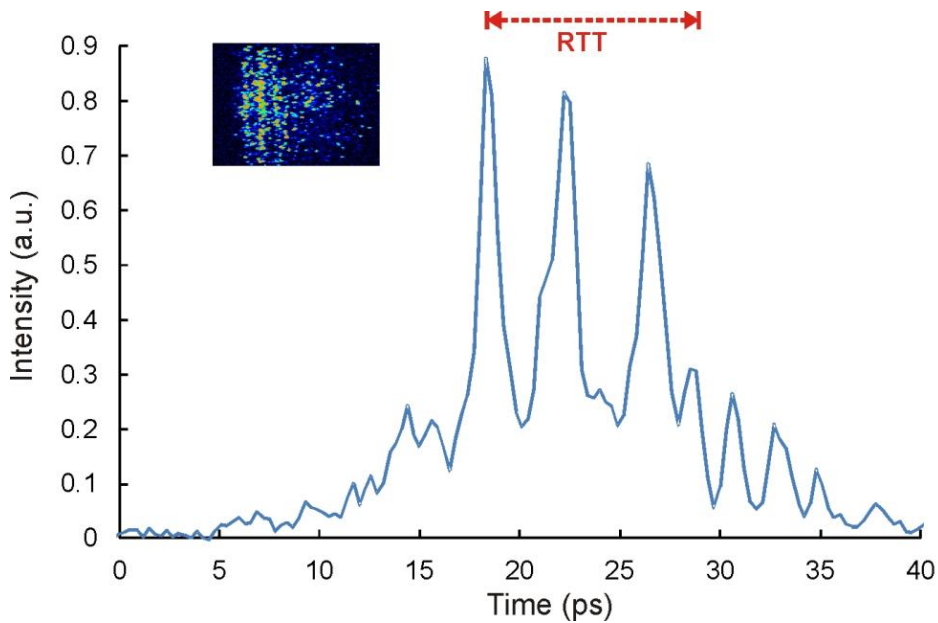

Figure 4. Photo from the streak camera screen and corresponding SR pulses. The round trip time (RTT) of the cavity related to lasing is shown by the red line.

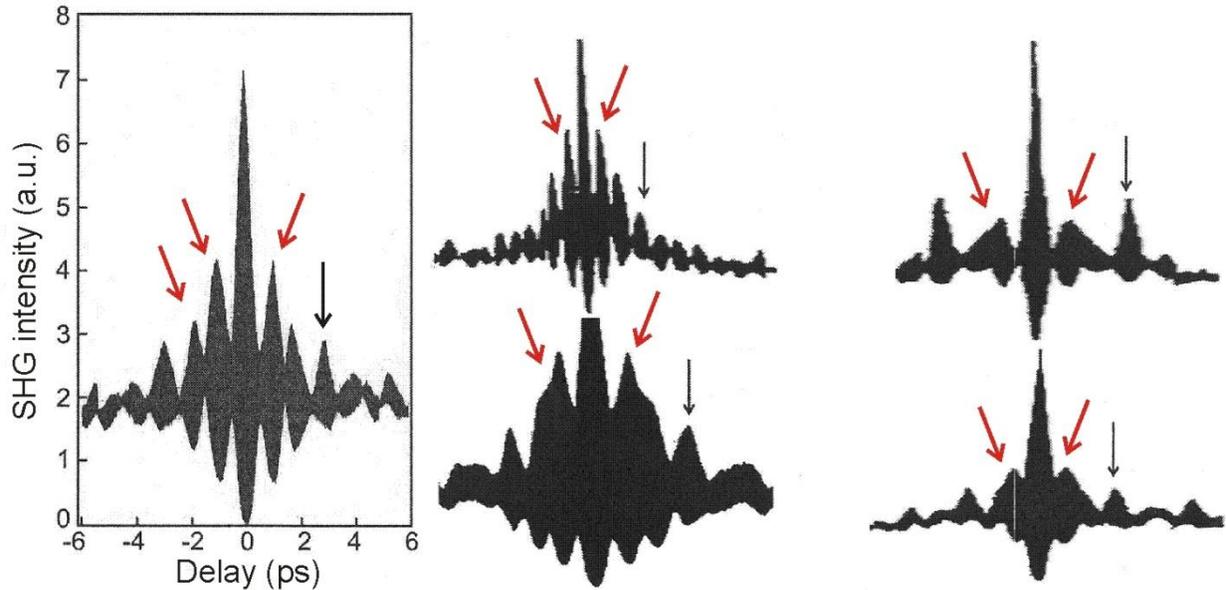

Figure 5. Autocorrelation traces of SR pulses at different values of *I* and *V*. The device length is 98 microns which corresponds to a round trip time of around 3 ps. The peaks at RTT are marked by the black arrows, the additional peaks caused by the population gratings are shown by red arrows.

Autocorrelation traces (Fig. 5) measured for SR pulses show additional structures or oscillations in addition to the structures of lasing. This can be explained by assuming that there is an oscillatory process linking the electromagnetic field and the e-h system typical of grating locking. The shape and number of coherent oscillations depend strongly on the driving conditions (both *I* and *V*) and the geometry of gain/absorber sections, unlike these for lasing which do not change significantly. SR emission travels along the longitudinal axis between the chip facets successfully being reabsorbed and re-emitted again in a similar way to Rabi-type oscillations. Multiple peaks and fringes at nonzero delays originate from the coherent interaction of the electromagnetic field with carrier density transient gratings. These peaks are marked by the red arrows and, as we see in Section III below, they never exist without the coherent population gratings.

The measurement of round trip times of lasing has been carried out to demonstrate the independence of these times on the driving conditions. A large number of measurements of over 20 devices of different lengths were performed, as shown in Figure 6. (It should be noted that the spectral and time domain measurements were consistent within the measurement error.) One can see that the change of the group refractive index for the Q-switching with respect to c.w. laser emission is less than 2 %, whilst it is typically within 3-23 % in case of SR, the largest measured values being around 35 %. Interestingly, the experimentally measured decrease of $n_g$ depends on the length of the structure, the maximum changes being for the shortest devices. The obtained results are consistent across the different semiconductor structures under test and the range of operating conditions. No increase of group refractive index is detected. Here, it should be pointed out that despite superluminal propagation effects, no real signal can be transmitted faster than the vacuum velocity of light, and the causality principle is not violated as Sommerfeld and Brillouin have previously demonstrated [27].

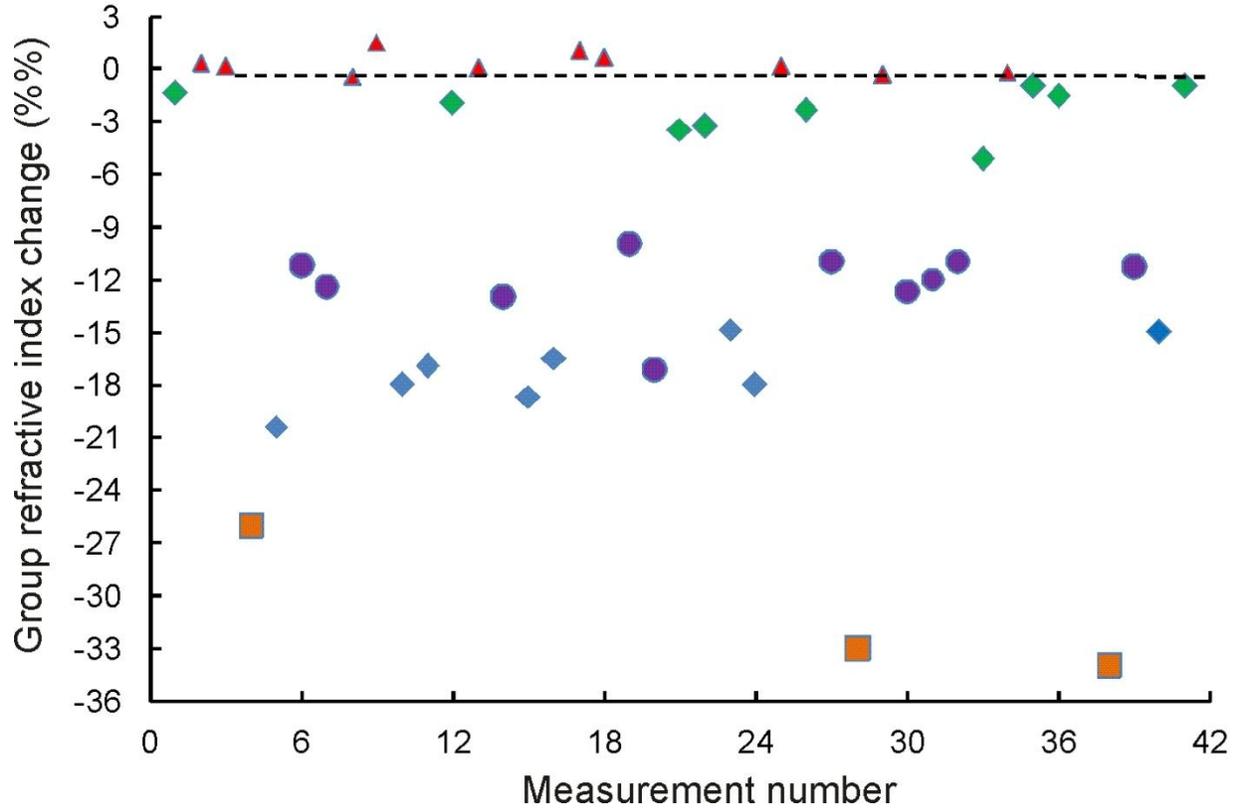

Figure 6. Variations of $n_g$ for different devices and different driving conditions. The initial values of the refractive index were calculated for lasing near the threshold. ▲ - Q-switching; ♦ - SR, 450 μm long cavity; ● – SR, 350 μm; ♦ - SR, 150 μm; ■ – SR, 100 μm.

### III. Theoretical modelling

The physical reason behind the 'fast light' phenomenon during the SR phase transition appears to be different from those faster-than-*c* propagation effects which have been recently observed in resonant media with population inversion [18-21]. Indeed, the semiconductor active medium in the GaAs/AlGaAs heterostructures studied does not exhibit any closely spaced gain or absorption lines with a strong anomalous dispersion region, which can be responsible for the fast light effect. However, by contrast to any laser, where the field is coherent while the active medium is incoherent, macroscopically large areas of the coherent macroscopic population exist in the semiconductor during the SR phase transition [22-25]. It has been previously suggested that the superluminal propagation of light in a coherent amplifier can occur as a results of the pulse reshaping [28-32].

Coherence of the semiconductor medium as well as induced coherent population gratings plays a decisive role in the phenomenon under study. Indeed, reflections of the electromagnetic field at the chip facets lead to co-existence of two counter propagating SR pulses. Due to coherence of the e-h system these two pulses each form a coherent transient grating with a spatial period equal to a half of the wave vector of the light $k$. Nonlinear interaction of the SR pulses with coherent gratings can result in variations of their group velocity. The impact of this mechanism on superluminal pulse propagation can be clarified by solving Maxwell-Bloch equations [9,33] with the explicit inclusion of the spatial carrier grating. The spatial distribution of the carrier density can be expressed as follows [9,34]

$$N(z,t) = N_0(z,t) + N_1(z,t)\cos 2kz. \quad (1)$$

The system of equations, which is governed the dynamic of the SR pulse generation, is given in dimensionless variables as [9,17,33]

$$\frac{\partial E^+}{\partial t} + \frac{\partial E^+}{\partial z} = P^+(1+i\delta) + i\kappa P^- \quad (2)$$

$$\frac{\partial E^-}{\partial t} - \frac{\partial E^-}{\partial z} = P^-(1+i\delta) + i\kappa P^+ \quad (3)$$

$$\frac{\partial P^+}{\partial t} = -\gamma_2 P^+ + E^+ N_0 + E^- N_1 + \Lambda^+{}_N \quad (4)$$

$$\frac{\partial P^-}{\partial t} = -\gamma_2 P^- + E^- N_0 + E^+ N_1 + \Lambda^-{}_N \quad (5)$$

$$\frac{\partial N_0}{\partial t} = -4(E^+ P^+ + E^- P^-) \quad (6)$$

$$\frac{\partial N_1}{\partial t} = -2(E^+ P^- + E^- P^+) \quad (7)$$

where $E^{\pm}$ and $P^{\pm}$ are the slowly varying amplitudes of the counter propagating electric fields and the active medium polarizations, respectively, $\gamma_2$ is the dimensionless polarization relaxation rate. Since we are interested in ultrafast phenomena taking place on a time scale of just a few picoseconds, we neglect the terms describing drive current, spontaneous emission, and carrier diffusion. The Eqs. (2) and (3) include explicitly the terms responsible for the interaction of the fields with the carrier grating as in the modelling of colliding pulse mode-locked semiconductor lasers where the carrier grating in the absorber plays a decisive role [26,35]. The strength of the interaction is determined by the coefficient $\kappa$ which is proportional to $N_1$ [35]. In contrast to SR generation by solid-state media and gases, there exists a strong relation between the refractive index and the carrier density in semiconductors. The parameter $\delta$ takes into account changes in the refractive index due to variations of $N$. $\delta$ is proportional to the linewidth enhancement factor [35]. Eqs. (6) and (7) include $\delta$–correlated Gaussian random functions $\Lambda^{\pm}_N$ describing spontaneous noise in a similar manner to that in our previous SR model [25]. The Eqs. (2)-(7) were solved with the initial and boundary conditions and the material parameters used in our previous SR models for GaAs/AlGaAs samples [36,37].

The initial value of $N(z,0)$ is 1 at the end amplifying sections and 0 at the central absorber and the field reflectivity of the chip facets is 0.57. Due to the random nature of the initialization process of the SR emission, individual distributions of $N_1(z,t)$ are also random. The coherence of the interaction of $E^{\pm}$ and $N(z,t)$ results in a built-up of negative values of both variables. Ultrafast alternations of the carrier density are accompanied by the ultrafast oscillations of the induced polarization of the medium $P^{\pm}(z,t)$. Figure 7 shows output SR pulses and oscillations of the e-h polarization during the generation of SR pulses with (blue line) and without (red line) the transient grating.

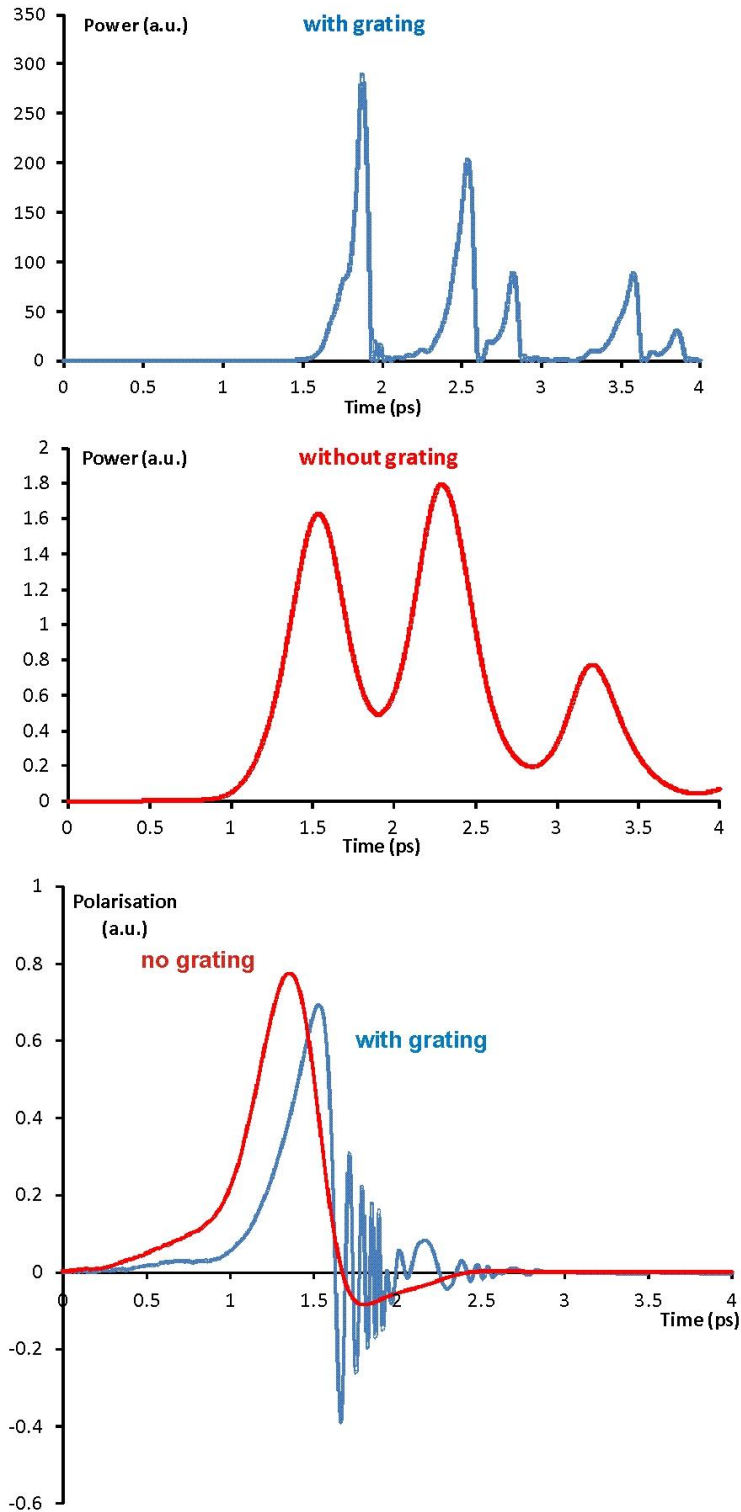

Figure 7. Calculated SR pulses and evolution of the polarization of the semiconductor medium during the SR emission generation. The sample length is 70 μm.

The coherent transient grating induced in the medium enhances strongly the polarization oscillations and result in the generation of shorter and more powerful SR pulses. These ultrafast oscillations lead to the emergence of additional peaks in the fringe-resolved SHG autocorrelations of SR pulses often observed experimentally (see, Fig. 5) [36, 37]. Figure 8 shows typical fringe resolved SHG traces of SR pulses generated from a 100 μm long structure calculated using the travelling wave model Eqs (1)-(7). Figure 8 presents the average of 40 individual SR realizations. The system of equations for 'no grating' case was solved for $\delta = \kappa = N_1 = 0$. It is clearly seen in Fig. 8 that the coherent transient grating leads to superluminal propagation of SR pulses. Indeed, the round trip time of the sample is 2.9 ps. If one takes the grating into account, the value reduces to 2.6 ps. The peaks corresponding to the RTT are shown by arrows in Fig. 8. The transient population gratings are responsible for the additional peaks within the RTT. The individual shape of the SHG peaks depends strongly on the small signal gain, initial value of the carrier density, the polarization relaxation time, and values of $\delta$ and $\kappa$. This model therefore demonstrates the importance of coherent gratings in the SR generation process in GaAs/AlGaAs semiconductor devices at room temperature.

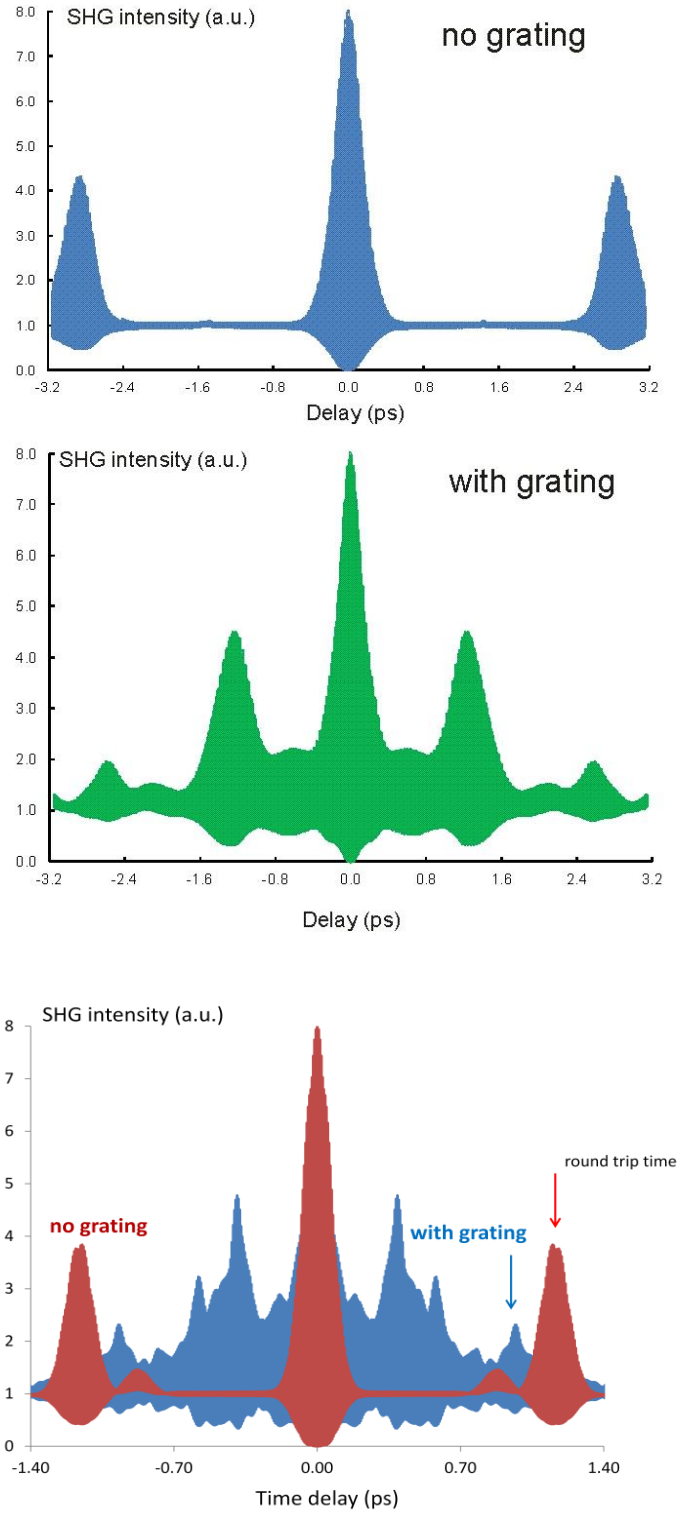

Figure 8. Calculated SHG autocorrelation traces of SR pulses. Compare with the experimental data shown in Fig. 5.

**IV. Discussion**

It has been previously established [22-25] that SR in semiconductors exhibits a large red shift of emission wavelength and occurs at much larger e-h concentrations than for conventional lasing. It should be noted here that there exist two additional effects due to the substantial nonlinear refractive index change in semiconductor heterostructures. The first one concerns with a decrease of the material refractive index with increasing wavelength (the chromatic dispersion). The second effect is a dependence of the refractive index of a semiconductor on the injected carrier density. These can in principle contribute to the measured variations of the group refractive index because both effects result in decreasing of the refractive index. Let estimate the impact of this effects.

The active layer of all tested structures is intrinsic bulk GaAs. The chromatic dispersion coefficient of it is $dn/d\lambda = -1.156 \mu m^{-1}$ in the range of from 800 to 900 nm. A typical red shift of the peak wavelength of SR emission is up to 20 nm with respect to the lasing wavelength [22]. This gives a change of the refractive index of about $-0.6$ %. Carrier-induced changes in the refractive index of group III-V semiconductors have been intensively studied both experimentally and theoretically (see for instance [38] and references therein). Three effects, bandfilling, band-gap shrinkage, and free-carrier absorption, can produce sizeable contributions to the total change in refractive index $\Delta n$. Typical carrier densities at which SR phase transition takes place in bulk GaAs are around $6 \times 10^{18}$ cm$^{-3}$, whereas the lasing threshold e-h concentration is around $(1.5-2.0) \times 10^{18}$ cm$^{-3}$ in the same structures [26]. As calculated in [38], this difference of the e-h density can lead to a total change of the refractive index of $\Delta n \sim -0.01$ which is less than 1%. Therefore, the chromatic dispersion and carrier-induced variations of the refractive index contribute together about 1% of the total value of the refractive index change. They cannot be the reason for a large decrease of the group refractive index observed experimentally.

The importance of coherent population gratings for the generation of SR emission in semiconductor structures has not been understood in previous studies [5-7, 22]. Here, we point out two issues which have been clarified during the present work. First, a sufficiently large value of coupling between oscillators and the radiation field is required for a system to exhibit an SR quantum phase transition at a certain temperature [12,13]. In our case of multiple section semiconductor structures the coupling is enhanced by 1) the existence of internal resonator and reflections of the radiation field at the cleaved chip facets and 2) the presence of optical gain at the excitonic part of the spectrum [25]. The emergence of population gratings makes light-matter coupling even more efficient and hence facilitates setting the SR phase transition up. The grating results in forward-backward coupling of counterpropagating SR pulses ensuring the establishment of mutual coherence of the electromagnetic field and the e-h system. The reflectivity of the facets of the GaAs/AlGaAs samples under test is much greater than the much lower reflection coefficients of the solid-state and gaseous samples in previous SR studies [3,4]. Therefore, the effect of forward-backward coupling of counter-propagating SR pulses and transient population gratings in semiconductor structures is much larger.

The second issue is concerned with self-organization from the homogeneous into a periodically patterned distribution in the e-h ensemble. It has been clearly understood recently that self-organization of emitters plays an essential role in Dicke quantum phase transition in Bose condensates coupled to a single mode of an optical cavity [14,39,40]. Cold-atom trapped Bose condensates experience self-organization and Dicke phase transition when they are pumped transversely by a laser. The characteristic $\lambda/2$ spatial periodicity has been predicted and experimentally observed. In the SR emission generation described here phase transition builds up within optical cavities formed by the cleaved facets. The e-h self-organization and the growth of carrier gratings are caused by two counterpropagating resonant internal electromagnetic fields at the excitonic part of the spectrum bouncing back and forth between the chip facets. The number of optical modes within a certain bandwidth depends obviously on the sample length. A typical optical bandwidth of the femtosecond SR pulses in our experiments is approximately 3-4 nm [22,24,25]. This means that there exist 3-4 or 15-20 modes within this bandwidth for 100 and 450 μm long samples, respectively. Each individual optical mode forms its own λ/2 population sub-grating. When more modes take part in the formation of the cumulative carrier grating, spatial variations of the e-h density are efficiently averaged out. This inevitably implies that the effect of transient coherent gratings is more pronounced in shorter samples where just a few optical modes contribute to the cumulative carrier grating. Figure 6 demonstrates that short (100-150 μm long) structures exhibit greater decrease of $n_g$. In addition, coherent beating of the electromagnetic field and optical doublets and triplets has been always observed in short samples and never in long (> 250 μm) ones [22,36,37]. These facts support the importance of the coherent population gratings in SR generation by semiconductor devices.

## V. Conclusion

In summary, this paper reports an experimental and theoretical study of the role of coherent transient gratings in causing superluminal pulse propagation during SR emission generation in GaAs/AlGaAs heterostructures. It is demonstrated that the group refractive index can be changed by about 30-40 % smaller during superradiant phase transition in semiconductors, much larger than that due to normal laser emission. The superluminal pulse propagation is attributed to the coherent pulse amplification in the semiconductor active medium and the formation of coherent transient gratings of the population inversion. The observed phenomenon gives definitive evidence that the SR emission generation is fundamentally different from normal lasing despite many similarities between SR multiple contact structures and Q-switched semiconductor lasers of similar semiconductor compositions.

The observed substantial decrease of the group refractive index is just another feature of the non-equilibrium macroscopically ordered BCS-like e-h condensed state formed in the semiconductor during the SR quantum phase transition [2,22,23]. We have reported the importance of coherent population gratings which have been neglected during the previous investigations of the SR emission generation in semiconductors. The inclusion of the transient grating effect allows for much better understanding of the SR quantum phase transition and allows a complete physical description.

The authors thank V.Olle for help and H.Kan and H.Ohta of Hamamatsu Photonics for the fabrication of the semiconductor structures.